\documentclass[conference]{IEEEtran}
\IEEEoverridecommandlockouts
\usepackage{cite}
\usepackage{amsmath,amssymb,amsfonts}
\usepackage{algorithm2e}
\usepackage{graphicx}
\usepackage{textcomp}
\usepackage{xcolor}
\usepackage{array}
\usepackage{multirow}
\usepackage{enumitem}
\usepackage{bbding}

\def\BibTeX{{\rm B\kern-.05em{\sc i\kern-.025em b}\kern-.08em
    T\kern-.1667em\lower.7ex\hbox{E}\kern-.125emX}}
\begin{document}

\title{TinySV: Speaker Verification in TinyML with On-device Learning\\
\thanks{Identify applicable funding agency here. If none, delete this.}
}

\author{\IEEEauthorblockN{Massimo Pavan\textsuperscript{\textsection}}
\IEEEauthorblockA{
\textit{Politecnico di Milano}\\
Milan, Italy \\
massimo.pavan@polimi.it}
\and
\IEEEauthorblockN{Gioele Mombelli\textsuperscript{\textsection}}
\IEEEauthorblockA{
\textit{Politecnico di Milano}\\
Milan, Italy \\
gioele.mombelli@mail.polimi.it}
\and
\IEEEauthorblockN{Francesco Sinacori}
\IEEEauthorblockA{
\textit{Infineon Technologies Italia s.r.l.}\\
Milan, Italy \\
francesco.sinacori@infineon.com}
\and
\IEEEauthorblockN{Manuel Roveri}
\IEEEauthorblockA{\textit{Politecnico di Milano}\\
Milan, Italy \\
manuel.roveri@polimi.it}
}

\maketitle

\begingroup\renewcommand\thefootnote{\textsection}
\footnotetext{These authors contributed equally to this work}

\begin{abstract}
TinyML is a novel area of machine learning that gained huge momentum in the last few years thanks to the ability to execute machine learning algorithms on tiny devices (such as Internet-of-Things or embedded systems). Interestingly, research in this area focused on the efficient execution of the inference phase of TinyML models on tiny devices, while very few solutions for on-device learning of TinyML models are available in the literature due to the relevant overhead introduced by the learning algorithms. 

The aim of this paper is to introduce a new type of adaptive TinyML solution that can be used in tasks, such as the presented \textit{Tiny Speaker Verification} (TinySV), that require to be tackled with an on-device learning algorithm. Achieving this goal required (i) reducing the memory and computational demand of TinyML learning algorithms, and (ii) designing a TinyML learning algorithm operating with few and possibly unlabelled training data. The proposed TinySV solution relies on a two-layer hierarchical TinyML solution comprising Keyword Spotting and Adaptive Speaker Verification module. We evaluated the effectiveness and efficiency of the proposed TinySV solution on a dataset collected expressly for the task and tested the proposed solution on a real-world IoT device (Infineon PSoC 62S2 Wi-Fi BT Pioneer Kit).

\end{abstract}


\section{Introduction}

Tiny Machine Learning (TinyML) recently became one of the most promising areas in the field of Machine Learning. By enabling machine and deep learning models and algorithms to operate on battery-operated devices\cite{warden_tinyml_2020, alippi_not_2017} (e.g., embedded and Internet-of-Things units), TinyML created a whole new class of tasks and applications ranging from Keyword Spotting (KS) \cite{warden_speech_2018}, i.e., recognizing a pre-determined word or command in an audio stream, to object or anomaly detection \cite{chowdhery_visual_2019, s23042344} in images or accelerometers data.%, anomaly detection with IMU sensors \cite{}, or even presence detection with radar \cite{pavan_tinyml_2022}. 

A growing literature exists in the field of TinyML \cite{david_tensorflow_2021, roveri2022tiny}. Solutions in this field aim at either designing efficient architectures for machine and deep learning models (e.g., neural networks models employing efficient and lightweight layers) \cite{howard_mobilenets_2017, tan2019efficientnet} or approximate computing strategies to optimize the memory and computational demand (e.g., quantization or pruning mechanisms) \cite{jacob_quantization_2017, liu_pruning_2020}. 

Interestingly, current solutions assume that the training phase of TinyML models is carried out in the Cloud where appropriate computing and memory resources are available, while just the inference phase is performed on the target tiny devices. 

Unfortunately, this approach does not allow TinyML solutions to exploit data collected directly from the field by the device, hence preventing the incremental training or adaptation of the TinyML algorithms during the operational life. Many applications that require on-device adaptation capabilities are consequently still not viable in TinyML. An example in this field is “Speaker verification” (SV) \cite{irum2019speaker}, a task that consists of recognizing the identity of a user by analyzing audio captions provided by the user as a reference and comparing them to newly collected audio data.
In this context, the implementation of a SV system on a tiny device would enforce relevant applications, including smart locks that can recognize their owners or smart objects offering different behaviors according to the specific person it is interacting with. 

In this work we propose, for the first time in the literature, the definition of \textit{Tiny Speaker Verification} (TinySV), a task specifically tailored to the \textit{on-device learning} context, and introduce a TinyML algorithm supporting the on-device learning of SV applications. The proposed solution has been specifically designed to:
\begin{itemize}
    \item Learn a TinyML model directly on-device in a \textbf{one-class} manner (with data belonging to only one class of label);
    \item Operate in a \textbf{few-shot} setting (hence enforcing the learning on a small amount of data);
    \item Run continuously in an ``always-on" manner on a tiny, battery-operated device. 
\end{itemize}

In more detail, the proposed solution operates in a \textit{text-dependant} way (i.e., a pre-determined keyword is used to recognize the identity of the speaker \cite{tu-youzhi2022}), and relies on a two-layer hierarchical solution comprising Keyword Spotting (KS) and Adaptive Speaker Verification (ASV) operating in a cascade manner.
The solution has been tested on a text-dependent SV dataset that has been expressly collected for this task, which is released to the community along with the code for the experiments and the implementation in the project repository\footnote{https://github.com/AI-Tech-Research-Lab/TinySV}.

The paper is organized as follows. Sec. \ref{sec:related-literature} introduces the related literature. Sec. \ref{task} formalizes the task of \textit{Tiny Speaker Verification} proposed in this work. The proposed solution is described in Sec. \ref{proposed_solution}. Sec. \ref{Experiments} describes the experimental settings and results. Details on the on-device implementation of TinySV are given in Sec. \ref{implementation}, while conclusions are finally drawn in Sec. \ref{conclusions}.

\section{Related Literature}
\label{sec:related-literature}

In this Section, we discuss the related literature in the following fields: TinyML (Section \ref{subsec: tinyml}), Incremental on-device Learning in TinyML (Section \ref{subsec: incremental}), and Speaker Verification (Section \ref{subsec: SV}).

\subsection{TinyML}
\label{subsec: tinyml}

TinyML is a field of study that combines embedded systems and machine learning (ML). It studies ML models and architectures designed to be executed on small and low-power devices, hence taking into account their severe technological constraints in terms of memory (less than $1$ MB of RAM available on-device), computation (clock frequency is in the order of hundreds of KHz), and power consumption (less than tens of mW) ~\cite{warden_tinyml_2020}. Most of the solutions present in this field focus on the design of \textit{approximated machine and deep learning solutions}~\cite{sanchez-iborra_tinyml-enabled_2020,david_tensorflow_2021}. In particular, techniques such as weight quantization\cite{jacob_quantization_2017}, pruning \cite{liu_pruning_2020} and gate-classification \cite{disabato_reducing_2018} have been developed to reduce the memory and computational demand of machine and deep learning models, while guaranteeing their accuracy \cite{ray_review_2021,alippi_moving_2018}. 

TinyML paved the way for a wide range of intelligent embedded applications like visual wake-word detection \cite{chowdhery_visual_2019}, anomaly detection with accelerometers \cite{s23042344}, and presence detection with radar \cite{pavan_tinyml_2022}.
Among the wide range of applications keyword spotting (KS) \cite{warden_speech_2018} received a lot of attention from both the academic and the industrial perspective thanks to the ability to detect the presence of a pre-determined word or command in a continuous audio stream. 

\subsection{Incremental on-device Learning in TinyML}
\label{subsec: incremental}

Incremental on-device TinyML is a novel and promising area of TinyML aiming to directly support the incremental learning of TinyML models on the tiny devices, hence overcoming the traditional ``train-on-cloud and deploy-on-device" paradigm in TinyML. 

Solutions present in the literature can be organized into two main categories \cite{rajapakse_intelligence_2022}: instance-based (called lazy learning) and model-based (called eager learning). 

\subsubsection{Instance-based}

The instance-based solutions present in the literature \cite{disabato_incremental_2020, disabato_tiny_2021} and \cite{rusci2023few} rely on a Convolutional Neural Network (CNN) to perform feature extraction and dimensionality reduction on the input data. In these models, the learning phase consists of storing the labeled representations, while the inference phase involves the computation of a distance metric between the unlabeled representation of the input data and the previously extracted representations. The main advantages of this approach lie in the fact that (i) the training, which is usually the most computationally demanding task in ML, consists just of storing a dimensionality-reduced version of the data and (ii) 
these solutions provide acceptable results even with a small amount of data available \cite{rusci2023few}.

\subsubsection{Model-based}

Model-based learning mostly relies on the use of an optimized version of backpropagation for the adaptation of neural networks directly on-device. All the solutions present in the literature freeze some parts of the neural network to reduce the number of weights that need to be trained \cite{lin_-device_2022, ramanathan_online_nodate}. The same approach is used in \cite{ren_tinyol_2021} on a task of anomaly detection. All these solutions rely on training in an online manner (i.e., train on one datum at a time and discard it), and for this reason, they are limited in their ability to learn complex patterns and exploit batches of data to avoid overfitting. %so as to avoid the need to store the data that a standard batch learning for multiple epochs would require. %the data to train over them for multiple epochs. 
A solution to enable learning over batches of data is explored in \cite{ravaglia_tinyml_2021}, which proposed to store only the latent representations (i.e., lighter representation of data in terms of memory occupation with respect to the complete datum) in order to perform multiple training epochs. Despite that, the amount of latent representations storable on tiny devices is usually orders of magnitude smaller than the one usually used in standard ML pipelines. For this reason \cite{pavan_tybox_2023} proposed a hybrid approach that continuously adapts the last layers of the network on batches of data stored as latent replays.
The only model-based solution present in the literature that does not rely on neural networks is \cite{sudharsan_train_2021}, an extremely efficient binary classifier that works on low-dimensional data.
We emphasize that all the model-based solutions present in the literature assume a large availability of labeled data to perform training, a requirement seldom satisfiable in the TinyML environment \cite{Warden_article_blog}. 

Currently, none of the works present in the On-device TinyML literature encompass on-device learning mechanisms able to work in a few-shot and one-class manner at the same time.

\subsection{Speaker Verification}
\label{subsec: SV}

The Speaker Verification (SV) task can be formalized as a binary classification problem where the goal, given an audio segment containing the voice of a user, is to distinguish whether this voice belongs or not to a previously enrolled speaker. The enrolled speaker is expected to provide a series of audio recordings containing his/her voice so as to configure the SV system. 

The SV task can be tackled with either a text-dependent approach (the user is expected to pronounce a pre-determined word to be recognized) or a text-independent one (the algorithm is expected to recognize the enrolled user independently from what they are saying) \cite{irum2019speaker}. 

Available solutions for SV include Gaussian Mixture- Model-Universal Background Models (GMM-UBM), Gaussian Mixture-Model Support Vector Machines (GMM-SVM), Joint Factor Analysis (JFA) and i-vectors \cite{fefactoranalysis} \cite{svoverview}. With the advent of deep learning and its strong representation and classification abilities, the research in SV took two different directions: deep learning models operating in traditional frameworks, e.g., the DNN/i-vector approach \cite{svoverview}, and sole deep learning models extracting a representation of speakers’ voice characteristics in a low-dimensional space called “embedding”, on which classification and comparison algorithms can run \cite{svoverview}.
Some works targeting low memory footprint applications are present in the literature \cite{Variani2014, heigold2016end, chen2015locally}. 
Among these articles, the ``d-vector-based method" introduced in \cite{Variani2014} is one of the most suitable ones for edge applications. This method relies on a neural network able to extract a voice-dependent low-dimensional vector, called ``d-vector", from input speech that can be used by an instance-based solution for recognizing the identities of the speakers. 

Interestingly, some reference datasets are present in the literature both for text-independent \cite{panayotov2015librispeech} and text-dependent SV \cite{qin2020hi, larcher2014text}, but all of them encompass long audio recordings ($>$ 3s), a fact that makes their usage harder while developing solutions for extremely constrained environments. 

All in all, none of the solutions for SV present in the literature is tailored for tiny devices nor presents a deployment on embedded devices encompassing both the enrollment and inference phases.

\section{Tiny Speaker Verification: the use case}
\label{task}

The goal of this section is to introduce TinySV, a new application for on-device learning and speaker verification in TinyML. 
We emphasize that the task is a particular type of text-dependent SV (i.e., recognizing the identity of the enrolled speaker from utterances of a specific word), in which both the keyword (i.e., the specific word or passphrase) and the identity of the speaker must be recognized at the same time from a continuous audio stream directly on a tiny device. 

In addition, this task must be tackled while keeping into consideration the relevant and challenging characteristics of the TinyML context:

\begin{itemize}
    \item the SV algorithm must be adapted directly on-device, meaning that a new user should be able to enroll in the SV application by providing examples of their voice directly through the target device;
    \item the algorithm must operate in a one-class manner, meaning that it should be able to learn to distinguish between the enrolled user and any other users only from data coming from the enrolled one;
    \item the algorithm must follow a few-shot learning approach, meaning that it should be able to operate even with few training data of the enrolled speaker; 
    \item the algorithm must match the strict technical requirements of tiny devices, meaning that it must operate requiring a small amount of memory and computation during both the inference and learning phase.
\end{itemize}

\begin{figure} [ht]
    \centering
    \includegraphics[width=0.5\textwidth]{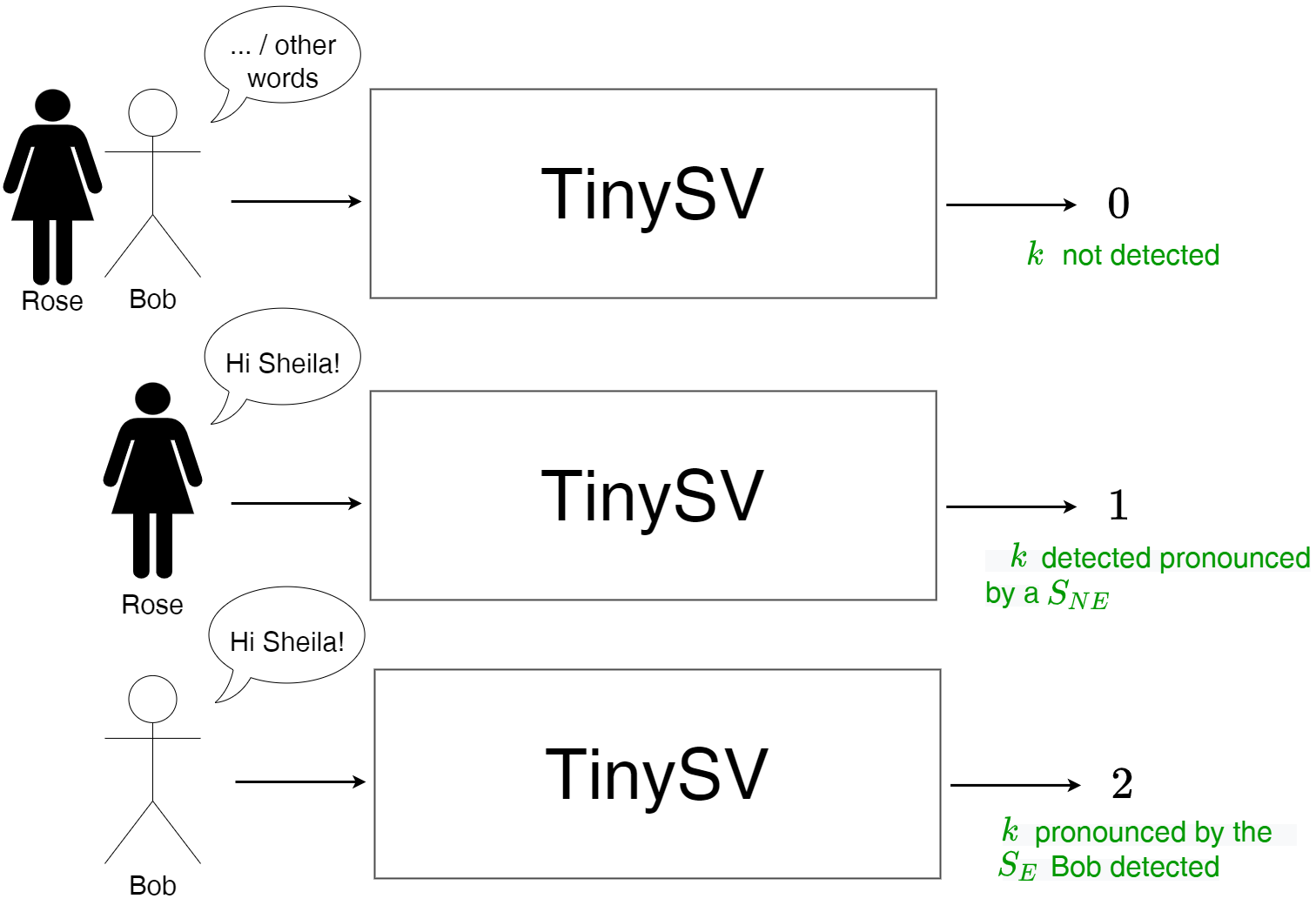}
    \caption{Examples of the use case, in which $k$ = "Sheila" and the enrolled speaker $S_E$ is Bob.}
    \label{fig:general_task}
\end{figure}

\begin{figure*} [t]
    \centering
    \includegraphics[width=0.75\textwidth]{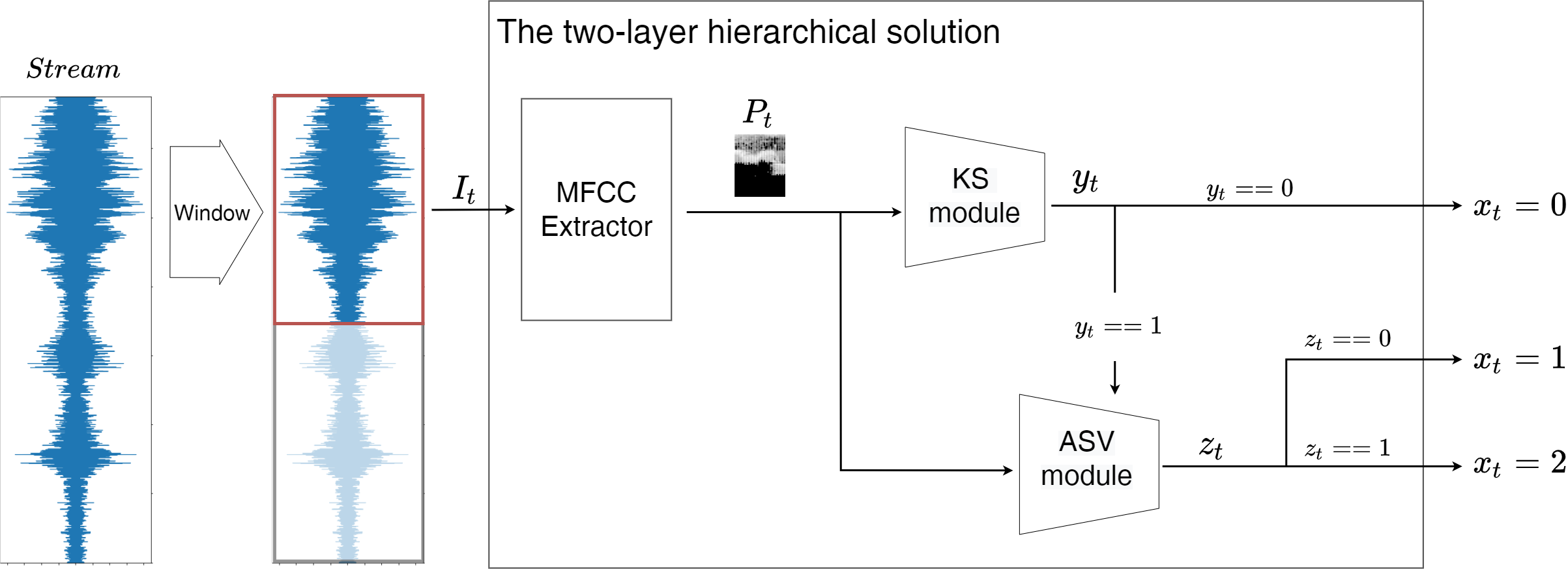}
    \caption{An high level representation of the proposed solution.}
    \label{fig:hi_level_arch}
\end{figure*}

More formally, the tiny device is continuously recording an audio stream by using a microphone characterized by the sampling frequency $f_r$. At time $t$, the most recent window $I_t$, whose length is $W$ seconds, is extracted from the stream and used as input for the algorithm. 

Given a pre-defined keyword $k$, the task of the TinySV algorithm is to assign a label $x_t \in \{0, 1, 2\}$ to the most recent segment $I_t$ of the stream where:

\begin{equation}
    x_t:
    \begin{cases}
        0: & k \text{ not present in } I_t  \\
        1: & k \text{ present in } I_t \text{ and pronounced by } S_{NE} \\ 
        2: & k \text{ present in } I_t \text{ and pronounced by } S_{E} \\
    \end{cases}
\end{equation}\\
where $S_E$ is the enrolled speaker (i.e., the speaker whose voice must be recognized by the algorithm), and $S_{NE}$ is any other, not-enrolled, speaker. The general use case of TinySV is depicted in Fig. \ref{fig:general_task}.

\section{Enabling TinySV: the proposed solution}
\label{proposed_solution}

The proposed solution for TinySV on audio streams relies on a two-layer hierarchical solution comprising: 

\begin{itemize}
    \item the Keyword Spotting (KS) module;
    \item the Adaptive Speaker Verification (ASV) module.
\end{itemize}

The KS model is used to determine if the audio segment $I_t$ under inspection includes the pre-determined keyword $k$.
If $k$ is detected in $I_t$, the audio segment is forwarded to the ASV module, which is meant to (i) create a personalized model for the enrolled speaker $S_E$ during the model adaptation phase and (ii) distinguish if $k$ was pronounced by $S_E$ or by a non-enrolled speaker $S_{NE}$ during the inference phase.  

We emphasize that the combination of the aforementioned two modules is used to address the problem formalized in Sec. \ref{task}, while a visual representation of the high-level pipeline of the proposed solution is depicted in Fig. \ref{fig:hi_level_arch}.

As detailed in what follows, before being used as input by the two modules, $I_t$ is pre-processed and transformed into a Mel-frequency cepstral coefficients (MFCC) spectrogram $P_t$ through a module called \textit{MFCC extractor}.
In order to reduce the number of operations needed to execute the pipeline on-device, the preprocessing is shared among the keyword spotting and the speaker verification module.

The rest of the section is organized as follows. In Sec. \ref{sub:preproc} the preprocessing phase performed by the MFCC extractor is described. The KS and ASV modules are described in Sec. \ref{kws-system} and \ref{speaker-verif}, respectively. Finally, a description of the two-layer hierarchical solution is drawn in Sec. \ref{sub:cascading}, followed by the comments on the memory requirements in Sec. \ref{sub:mem}.

\subsection{MFCC Extractor}
\label{sub:preproc}

The goal of this module is to transform the raw input $I_t$ into an MFCC spectrogram $P_t \in \mathbb{R}^{i \times j}$, highlighting the relevant audio features present in the data and, at the same time, reducing the data dimensions. 

The MFCC extractor relies on the pre-processing pipeline used in \cite{zhang2017hello} for keyword spotting, receiving in input a $W$-second long audio record sampled at $f_S$ (hence represented by a vector of dimension $W \cdot f_S$), and producing as output a $i \times j$ Mel Frequency Cepstral Coefficients (MFCC) spectrogram, being $i$ the number of frequency bins extracted from the pre-processing pipeline and $j$ the number of audio segments obtainable from a single window. The MFCC extractor operates by splitting the $W$-second long input into $\lambda$-seconds long audio segments and processing them through the use of FFT and Mel frequency downsampling. Since the $\lambda$ second-long segments are overlapped with a stride value of $\phi$, the value of $j$ can be computed as $j = W/\phi - \lambda/\phi$. 

In the proposed implementation and experimental section, the input $I_t$ (characterized by $W=1$s, $f_r=16$KHz) is preprocessed into a spectrogram $P_t$ of dimensions $i = 40 \times j = 49$, while $\lambda$ is equal to $30$ms and $\phi = 20$ms.

\subsection{The KS module}
\label{kws-system}

The KS module aims at recognizing if $I_t$ contains the pre-determined keyword $k$. The problem can be formalized as a binary classification task, whose goal is the association of $I_t$ to a label $y_t \in \{0, 1\}$ where:

\begin{equation}
    y_t:
    \begin{cases}
        0: & k \text{ not present in } I_t \\
        1: & k \text{ present in } I_t \\
    \end{cases}
    \label{eq:task_KS} .
\end{equation}\\

The KS module consists of a Convolutional Neural Network (CNN) $\Phi_k$ trained in a supervised manner to distinguish among silence, unknown words (i.e., speech that does not contain the keyword $k$), and the keyword $k$. It receives in input the spectrogram $P_t$ and, following other architectures used for keyword spotting\cite{sainath2015convolutional}, produces as output one of the 3 classes (i.e., silence, unknown, and keyword). The assigned value is $y_t = 0$ in the case in which the network assigns the silence or unknown class to the datum, $y_t =1 $ if it recognizes a keyword. 

\begin{figure}[ht]
    \centering
    \includegraphics[width=0.5\textwidth]{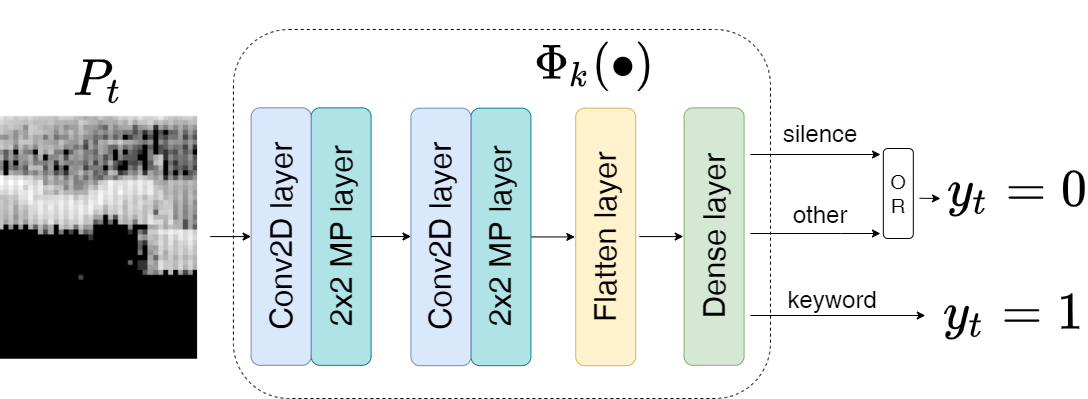}
    \caption{The architecture of the neural network used for keyword spotting.}
    \label{fig:KS_arch}
\end{figure}

$\Phi_k$ is organized as the state-of-the-art architecture labeled as \textit{cnn-trad-fpool3} proposed in \cite{sainath2015convolutional}, consisting of two 2D-convolutional/max-pooling blocks, comprising a 2D convolutional layer (characterized by a number $m$ of $r \times q$ filters and stride $=s$) and a $2 \times 2$ 2D max pooling layer, a flattening layer and a dense layer (characterized by a number $a$ of neurons). A high-level representation of the architecture is depicted in Fig. \ref{fig:KS_arch}.

$\Phi_k$ is also characterized by its total number of weights $\omega_{\Phi_k}$ and by the number of parameters required to store its activation $\alpha_{\Phi_k}$, which can be estimated as:

\begin{align*}
\begin{split}
\omega_{\Phi_k} = \sum_{l \in \Phi_k} \omega_l, \\
    \alpha_{\Phi_k} = \sum_{l \in \Phi_k} \alpha_l .
\end{split}
\end{align*}
being $\omega_l$ and $\alpha_l$ the number of weights and the output dimension of a layer $l$, respectively. 
$\omega_{\Phi_k}$ and $\alpha_{\Phi_k}$ obviously depend on the hyperparameters of the specific implementation of $\Phi_k$. The hyperparameters and the $\alpha_l$ and $\omega_l$ of the processing layers in $\Phi_k$ used for the on-device implementation in Sec. \ref{implementation} are reported in Tab. \ref{table:KS_CNN}. 

\begin{table}[ht]
\begin{center}
\begin{tabular}{|c|c|c|c|}
\hline
$l$ & Hyperparameters & $\alpha$ & $\omega$ \\ \hline \hline
                        
Input     & -                              & 1960 & 0  \\ \hline
Conv2D    &  $r=8$, $q=20$, $m=16$, $s=2$  & 8000 & 2576 \\ \hline
MP 2D &  $2\times2$                    & 1920 & 0 \\ \hline
Conv2D    &  $r=4$, $q=10$, $m=32$, $s=1$  & 3840 & 20512 \\ \hline
MP 2D &  $2\times2$                    & 960 & 0 \\ \hline
Flatten   &  -                             & 960 & 0 \\ \hline
Dense     & $a = 3$                        & 3 & 2883 \\ \hline \hline
Tot. $\Phi_k$ & & 17,643 & 25,971 \\ \hline
\end{tabular}
\end{center}
\caption{Hyperparameters, $\alpha$ and $\omega$ values of the $\Phi_k(\bullet)$ used in the on-device implementation.}
\label{table:KS_CNN}
\end{table}

\subsection{The ASV module}
\label{speaker-verif}

The task of the ASV module is to recognize if the keyword $k$ contained in the audio record $I_t$ was pronounced by the enrolled speaker $S_E$ or by another, non-enrolled, speaker $S_{NE}$. As before, the problem can be formalized as a binary classification task that consists of associating to $I_t$ a label $z_t \in \{0, 1\}$ where:

\begin{equation}
    z_t:
    \begin{cases}
        0: & k \text{ was pronounced by } S_{NE} \\
        1: & k \text{ was pronounced by } S_E \\
    \end{cases}
    \label{eq:task_SV} .
\end{equation}\\

The ASV module consists of a fixed d-vector extractor model $\Phi_f(\bullet)$ and an adaptive instance-based model used for the classification, $\Phi_c(\bullet)$. Both models are now detailed.

\subsubsection{The convolutional d-vector extractor $\Phi_f(\bullet)$}

The generated spectrograms $P_t$ are used as inputs for a convolutional neural network $\Phi_f(\bullet)$. 
Following a transfer learning approach $\Phi_f(\bullet)$ is developed by training a neural network to perform a speaker classification task in a supervised manner, and then removing its final classification layers. In more detail, $\Phi_f(\bullet)$ is composed of a batch normalization layer, a sequence of 2D convolution (characterized by a number $m$ of $ r \times q$ filters and stride = $s$) and Maxpooling layers, and a final flattening layer. A high level representation of the $\Phi_f(\bullet)$ architecture is provided in Fig. \ref{fig:SV_arch}. 

\begin{figure}[ht]
    \centering
    \includegraphics[width=0.5\textwidth]{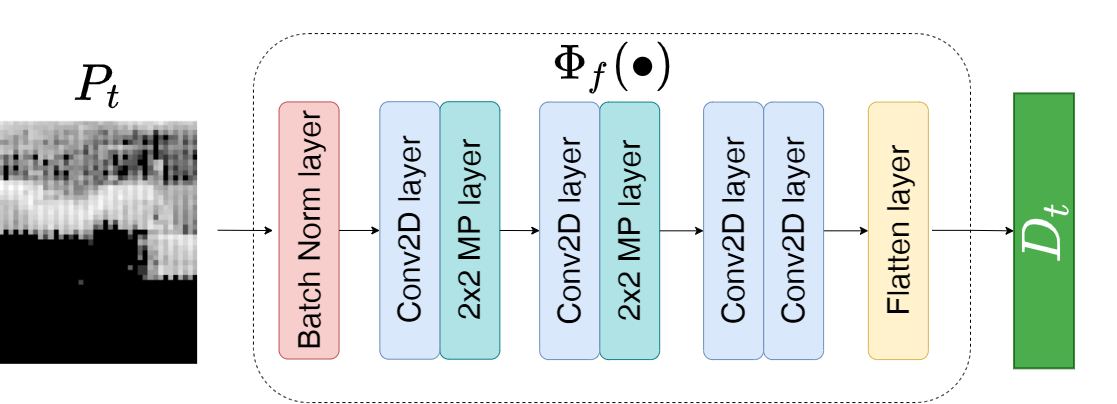}
    \caption{The architecture of the neural network used for extracting the d-vectors.}
    \label{fig:SV_arch}
\end{figure}

$\Phi_f(\bullet)$ is characterized by its total number of weights $\omega_{\Phi_f}$ and by the number of parameters required to store its activation $\alpha_{\Phi_f}$, which, similarly to $\omega_{\Phi_k}$ and $\alpha_{\Phi_k}$, can be estimated as:

\begin{align*}
\begin{split}
\omega_{\Phi_f} = \sum_{l \in \Phi_f} \omega_l, \\
    \alpha_{\Phi_f} = \sum_{l \in \Phi_f} \alpha_l .
\end{split}
\end{align*}
being $\omega_l$ and $\alpha_l$ the number of weights and activations of a layer $l$ of $\Phi_f$, respectively. 
The hyperparameters and values of the $\alpha_l$ and $\omega_l$ of the layers in $\Phi_f$ used for the experiments in Sect. \ref{Experiments} and in the on-device implementation in Sec. \ref{implementation} are reported in Tab. \ref{table:DvE_CNN}.  

\begin{table}[ht]
\begin{center}
\begin{tabular}{|c|c|c|c|}
\hline
$l$ & Hyperparameters & $\alpha$ & $\omega$ \\ \hline \hline
                        
Input     & -                              & 1960  & 0          \\ \hline
BatchNorm     & -                              & 1960  & 4          \\ \hline
Conv2D    &  $r=3$, $q=3$, $m=8$, $s=1$  & 15680 & 80  \\ \hline
MP 2D &  $3\times3$                    & 1664 & 0     \\ \hline
Conv2D    &  $r=3$, $q=3$, $m=16$, $s=1$  & 3328 & 1168  \\ \hline
MP 2D &  $2\times2$                    & 768   & 0 \\ \hline
Conv2D    &  $r=3$, $q=3$, $m=32$, $s=1$  & 384 & 4640  \\ \hline
Conv2D    &  $r=3$, $q=3$, $m=64$, $s=2$  & 256 & 18496  \\ \hline
Flatten   &  -                             & 256          & 0  \\ \hline \hline
Tot. $\Phi_f$ & & 26,656 & 24,388 \\ \hline
\end{tabular}
\end{center}
\caption{Hyperparameters, $\alpha$ and $\omega$ values of the $\Phi_f(\bullet)$ used in the on-device implementation.}
\label{table:DvE_CNN}
\end{table}

The latent representation $D_t \in \mathbf{R}^{d}$ (where $d$ correspond to the value $\alpha$ of the Flatten layer of $\Phi_f$) that $\Phi_f(\bullet)$ produces in output is called the D-vector, and it will be used as input for the training and inference of the classification model $\Phi_c(\bullet)$. In the experiments and in the on-device implementation, $d = 256$. 

\subsubsection{The instance-based model $\Phi_c(\bullet)$}
\label{sec:classification-algs}

It is the only part of the pipeline that is adapted directly on-device. It operates in two distinct phases: \textit{the learning phase} and \textit{the inference phase}.

\begin{figure}[ht]
    \centering
    \includegraphics[width=0.5\textwidth]{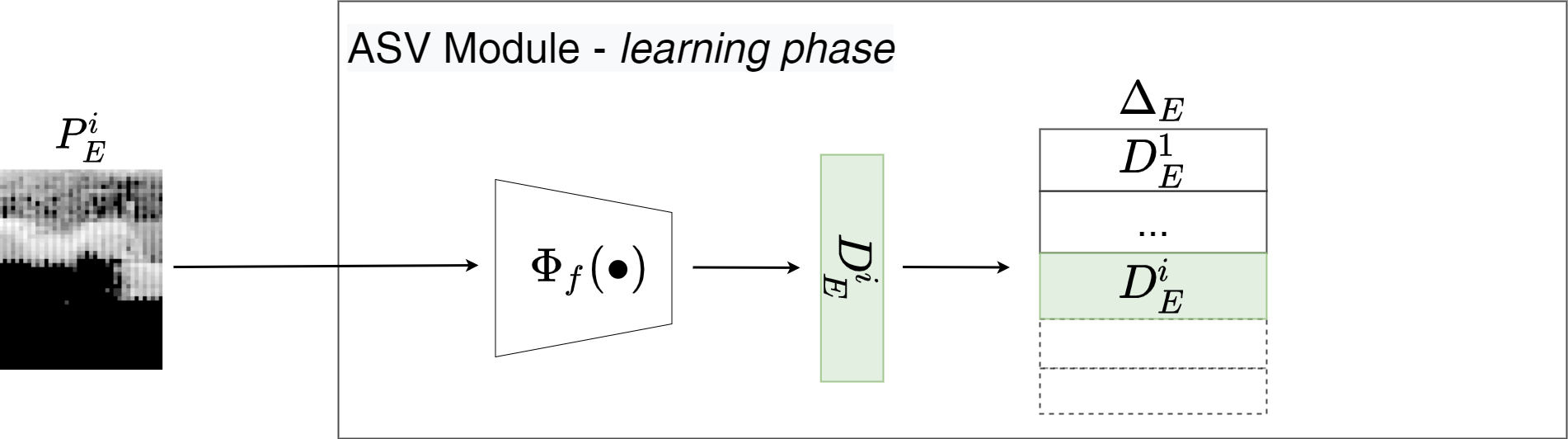}
    \caption{The adaptation phase of the proposed adaptive speaker verification model.}
    \label{fig:fe-cnn-scheme}
\end{figure}

\textit{2a) Learning phase:}
Being $\Phi_c(\bullet)$ an instance-based model, the training phase of the algorithm consists just in the collection of a pre-determined number $n$ of enrollment D-vectors $D_E$, collected from the enrolled Speaker $S_E$. This set of D-vectors is called \textit{the enrollment set} $\Delta_E = \{ D_E^1, ..., D_E^i ..., D_E^n\}$, being $D_E^i$ the i-th D-vector generated from the $i$-th Spectrogram $P_E^i$ that contains the keyword $k$. 
The Learning phase is depicted in Fig. \ref{fig:fe-cnn-scheme}. In the on-device implementation described in Sect. \ref{implementation}, the value $n = 16$ was used, while different values of $n$ were tested in the experiments.

\textit{2b) Inference phase:}
During the inference phase, the cosine similarity between the newly collected D-vector $D_t$ extracted from $\Phi_f(\bullet)$ and all the other vectors in $\Delta_E$ is computed and the best-match cosine similarity $\sigma(\bullet)$, defined as follows, is computed:

\begin{align}
    \sigma (D_t, \Delta_E) = \max_{\{D_i \in \Delta_E\}} \frac{D_t \cdot D_{i}} {\lvert \lvert D_t \rvert \rvert \cdot\lvert \lvert D_{i} \rvert \rvert} 
\end{align}

This value is compared to a user-defined threshold $\tau$ that can be tuned by the user in order to control the false positive vs false negative trade-off.  
Formally, the class $z_t$ is assigned to $D_T$ by using the formula:

\begin{align}
    z_t = \begin{cases}
        1 \; \text{ if } \; \sigma > \tau      \\
        0 \; \text{ if } \; \sigma \leq \tau   
    \end{cases}\ .
    \label{formula:zt}
\end{align} 

We emphasize that during inference phase, this approach requires having enough memory to keep the entire set of enrollment D-vectors $\Delta_E$ stored, togheter with the memory to store the input D-vector $D_t$. This aspect is deepened in Sect. \ref{sub:mem}.
The Inference phase is depicted in Fig. \ref{fig:fe-cnn-inference}.

\begin{figure}[ht]
    \centering
    \includegraphics[width=0.5\textwidth]{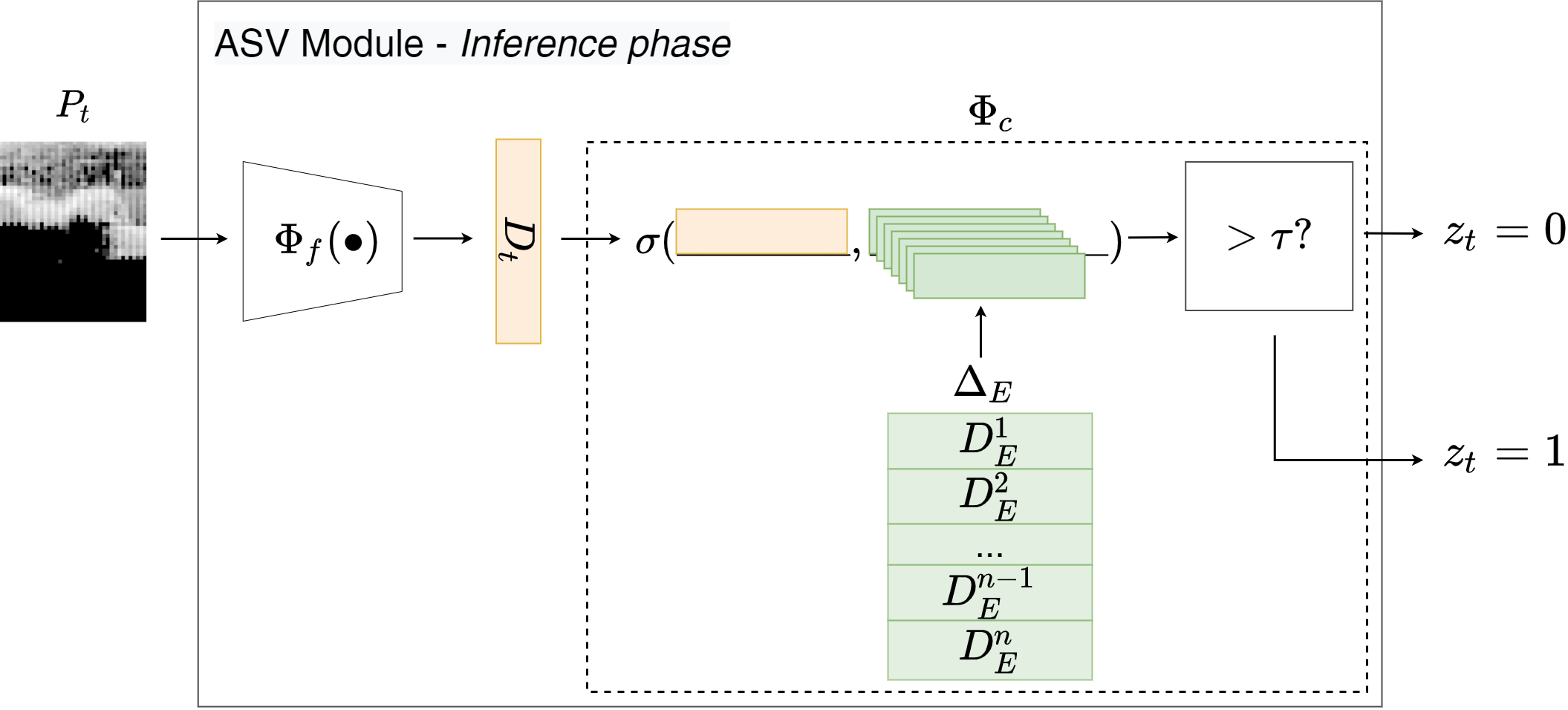}
    \caption{The inference phase of the proposed adaptive speaker verification model.}
    \label{fig:fe-cnn-inference}
\end{figure}

\subsection{The two-layer hierarchical solution}
\label{sub:cascading}
By executing the two proposed modules in a hierarchical manner, it is possible to enable the execution of TinySV on a tiny device. The pseudo-code provided in Alg. \ref{alg:two} describes the execution of the proposed two-layer hierarchical solution algorithm.

\begin{algorithm}
\caption{Pseudocode of the proposed two-layer hierarchical solution}\label{alg:two}
\textbf{Input:} $I_t$\\
\textbf{Output:} $x_t \in \{0, 1, 2\}$ \\

$P_t \gets MFCC(I_t)$\;
$y_t \gets \Phi_k(P_t)$\;
  \eIf{$y_t == 1$}{
    $D_t\gets \Phi_f(P_t)$\;
      \eIf{$|\Delta_E| < n$}{
        $\Delta_E \gets \Delta_E \cup D_t$\;
      }{
      $y_t \gets \Phi_c(D_t) $\;
      \eIf{$z_t == 0$}{
      $x_t \gets 2$\;
      }{$x_t \gets 1$\;}
  }
  }{$x_t \gets 0$\;
    }
\end{algorithm}

We enforce that, since the two algorithms are executed in a hierarchical fashion, the ASV module is executed only when the keyword $k$ is detected by the KS module. In this sense, the KS module acts as a filter, almost halving the amount of computation that would be performed at each inference cycle if the two algorithms were being executed in parallel, and it ensures the quality of the data given as input to the ASV module by centering the window of the input data on the keyword. 

\subsection{Memory requirements}
\label{sub:mem}

The memory requirements for each component of TinySV, i.e., the intermediate computations $I_t$, $P_t$, and $D_t$ and the models $\Phi_k$, $\Phi_f$, and $\Phi_c$, can be estimated with the formulas provided in Tab. \ref{parameters-composition}. 
We highlight that this estimation is system-agnostic, and thus does not consider any form of on-device optimization of a specific toolchain for the neural networks.

We emphasize that the memory of all the components can be computed as the product of the number of parameters required by the component and the precision $b$ (e.g., 1 Byte, 4 Bytes ...) in which they are stored.
In Tab. \ref{parameters-composition} the memory requirements of the components implemented in the on-device implementation in Sec. \ref{implementation} are also reported. For this estimation, the value $b=4$B was considered for all the components except for $I_t$, which is stored with a $b_1=2$B precision.

\begin{table}[ht]
\begin{center}
\begin{tabular}{|ccc|}
\hline
Component & Estimation formula & On-device mem. estimation \\ \hline \hline
$I_t$ & $(f_r \times W) \times b_1$ & 32 kB \\ \hline 
$P_t$ & $(i \times j) \times b$ & 7.5 kB \\ \hline
$D_t$ & $d \times b$ & 1 kB \\ \hline \hline
$\Phi_k$ - weights & $ \omega_{\Phi_k} \times b $ & 101.44 kB \\ 
$\Phi_k$ - act. & $ \alpha_{\Phi_k}  \times b $ & 68.92 kB \\ \hline
$\Phi_f$ - weights & $ \omega_{\Phi_f} \times b $ & 95.27 kB\\ 
$\Phi_f$ - act. & $  \alpha_{\Phi_f}  \times b $ & 104.12 kB \\ \hline
$\Phi_c$ & $(d \times n) \times b$ & 16 kB \\ \hline
\end{tabular}
\end{center}
\caption{Memory estimation for each component of TinySV.}
\label{parameters-composition}
\end{table}

\section{Experimental Setting and Results}
\label{Experiments}

In this section, we describe the experiments performed to analyze the performance of the ASV module. The experimental setting is outlined in Sect. \ref{ASV}. In Sect. \ref{subsect:comparisons} the two proposed comparison are detailed, while in Sect. \ref{subsect:results} the experimental results are provided.

\subsection{Experimental Setting}
\label{ASV}

The experimental setting for the ASV module was designed keeping in mind the one-class, few-shot conditions described in Sect. \ref{task}.
The one-class condition has been ensured by enrolling one speaker at a time and using only samples from that speaker to perform the enrollment. The few-shot conditions have been tested by limiting the number of samples $n$ used for the training phase. We provide the results for different values of $n$, i.e., $n = \{1, 8, 16, 64\}$.

\subsubsection{The collected dataset}

For the test of the proposed ASV model, we used a newly collected dataset comprising 376 recordings of the locution "Hey Cypress" pronounced by 4 different speakers (3 Male subjects and 1 Female, 94 recordings per subject). The mother tongue of all the speakers is Italian, so a possible bias in the English accent is present in the dataset. Training (68\%), validation (16\%) and test (16\%) sets have been extracted from the dataset for each user. 

The length of the recordings in the dataset is 1 second, compatible with the length of the proposed time window $W$. It is worth noting that a manual alignment of such samples has been performed to center the "Hey Cypress" phrase in the middle of the 1-second audio window.

\subsubsection{The ASV module}

In the ASV module used in the experiments, the implementation of $\Phi_f$ described in Tab. \ref{table:DvE_CNN} was obtained
from a model originally trained for a speaker classification task on the LibriSpeech-train-100 dataset \cite{panayotov2015librispeech}. Further details on the training of $\Phi_f$ can be found in the project repository.
$\Phi_c$ has been evaluated by considering each combination of the enrolled speaker $S_E$ and the number $n$ of d-vectors used to build the model. The values that have been tested for the parameter $n$ are $\{1, 8, 16, 64\}$, while all the four speakers in the dataset were used one at a time as $S_E$. 

\subsubsection{Metrics and evaluation}

Four different metrics were selected for the evaluation of the proposed solution: accuracy, F1 score, Equal Error Rate (EER), and Area Under Curve (AUC). The first two figures of merit evaluate the performance of the algorithm on the testing set after the setting of the parameter $\tau$, while the last ones are independent from that parameter and are computed on the validation set. 

In order to compute the accuracy and F1 score results for each speaker, the tunable parameter $\tau$ was set to %$Th_{EER}^{S}$, i.e. 
the threshold value corresponding to the Equal Error Rate for the speaker $S$ computed on the validation set. 

For all the figure of merit and values of $n$, we provide the average results of the 4 models of the speakers included in the dataset. 

\subsection{The proposed comparisons}
\label{subsect:comparisons}

As a comparison for the ASV module, we considered the following two solutions coming from the SV literature:

\subsubsection{Mean Cosine Similarity (MCS)}
\label{sec:meancos}

This solution maintains the same d-vector extractor $\Phi_f(\bullet)$ used in the proposed ASV module, but replacing the similarity metric $\sigma(\bullet)$ with the \textit{mean cosine similarity}. This metric is common in the Speaker Verification literature, and it consists in computing the cosine similarity between $D_t$ and $D_{AVG}$, extracted from $\Delta_E$ by computing the element-wise average of the d-vectors in the set. 
The memory requirements of this model are equal to $d \times b$, and, differently from the ones of the proposed $\Phi_c$, it does not vary with $n$.

\subsubsection{GE2E LSTM}

To provide a comparison with a state-of-the-art system, we tested an implementation of the Speaker Verification algorithm described in \cite{8462665} and \cite{yistlin}.
Similarly to our ASV module, this solution encompasses a d-vector extractor $\Phi_f^{LSTM}$ and a similarity metric.
$\Phi_f^{LSTM}$ is an LSTM neural network, with three layers, each containing 256 nodes. The network was trained with a \textit{generalized end-to-end} loss that aims at training models that better emphasize the differences in the feature space. The similarity metric used in this work is the \textit{Mean Cosine Similarity} described in the other comparison. 
This solution is not meant to be run on tiny devices, since $\Phi_f^{LSTM}$ requires more than 4MB only for storing the weights.

Technical details on the implementation of the two comparisons can be found in the project repository.

\subsection{Experimental Results}
\label{subsect:results}

The results of the proposed solution and of the comparison on the accuracy, F1 score. EER and AUC metrics are provided in Tab. \ref{tab:res_acc_F1}.

\begin{table}[ht]
\begin{center}
\begin{tabular}{|m{1cm}cccccm{0.8cm}|}
\hline
solution & metric & $n=$ 1 & 8 & 16 & 64 & Tiny device \\ \hline \hline

\multirow{4}{1.5cm}{ASV(our)} & Acc. & 0.773 & 0.825 & 0.833 & 0.846 & \multirow{4}{1.5cm}{\Checkmark} \\
 & F1 & 0.639 & 0.708 & 0.732 & 0.739 & \\
  & EER & 0.244 & 0.099 & 0.058 & 0.038 & \\
 & AUC & 0.855 & 0.953 & 0.975 & 0.987 & \\ \hline \hline
\multirow{4}{1.5cm}{MCS} & Acc. & 0.773 & 0.816 & 0.825 & 0.770 & \multirow{4}{1.5cm}{\Checkmark} \\  
 & F1 & 0.639 & 0.703 & 0.725 & 0.725 & \\ 
 & EER & 0.244 & 0.138 & 0.157 & 0.160 & \\  
 & AUC & 0.855 & 0.883 & 0.895 & 0.879 & \\ \hline \hline
\multirow{4}{1.5cm}{GE2E\cite{8462665}} & Acc. & 0.883 & 0.937 & 0.966 & 0.975 & \multirow{4}{1.5cm}{\XSolidBrush} \\  
 & F1 & 0.815 &  0.876 & 0.932 & 0.946 &\\
 & EER & 0.100 & 0.037 & 0.020 & 0 & \\  
 & AUC & 0.892 & 0.985 & 0.997 & 1 &
 \\ \hline 

\end{tabular}
\end{center}
\caption{Comparison between our ASV module and the comparisons.}
\label{tab:res_acc_F1}
\end{table}

The results show that our solution is extremely competitive with respect to the state-of-the-art solution meant to be run on larger, more flexible devices, while at the same time improving the state-of-the-art approach for tiny devices. 

Indeed, in all the metrics, the proposed solution outperforms the \textit{MCS} approach, particularly in the threshold-independent metrics EER and AUC, and with larger values of $n$. As expected, the \textit{MCS} and ASV approaches are equivalent and have exactly the same performance in the case $n=1$. Interestingly, the \textit{MCS} approach reported the worst performance with $n=64$, indicating that this type of model fatigues in incorporating the knowledge from larger, noisy enrollment datasets. The great differences in the EER and AUC metrics (i.e., 8\% - 10\%) between the proposed ASV and \textit{MCS} indicate also that with the proposed \textit{Best-Match Cosine Similarity} better tradeoffs are possible in the selection of the parameter $\tau$.

Compared to the \textit{GE2E LSTM} approach, the proposed ASV approach has a reduction in performance in the order of 2\% - 4\% for threshold-independent metrics, and in the order of 10\% - 20 \% in the threshold-dependent metrics. The proposed solution is nevertheless at least an order of magnitude less memory-demanding, and thus can be executed on tiny devices. 

\section{On-device implementation}
\label{implementation}

The proposed TinySV solution has been implemented on an off-the-shelf hardware platform to test its performance in a real-world scenario. 
The aim of this section is to describe the on-device implementation of TinySV, in which both the enrollment phase and the inference phase are executed on the target device. 

At startup, the TinySV demo application asks the user to provide the enrollment samples by pronouncing $n=16$ times the keyword $k$ = ``Sheila". Afterwards, the model switches to the \textit{inference phase} and recognizes if $k$ was pronounced by the enrollment user $S_E$ or not.

A video of the demo application can be found in the project repository, and a frame of the video is presented in Fig. \ref{fig:frame}.

The section is organized as follows. In Sec. \ref{sub:board} the considered hardware platform is presented. In Sec. \ref{sub:impl-details} the implementation details are reported, while Sec. \ref{sub:measured-mem} reports all the considerations on the measured memory occupations, power consumption and execution times.

\begin{figure}
    \centering
    \includegraphics[width=0.48\textwidth]{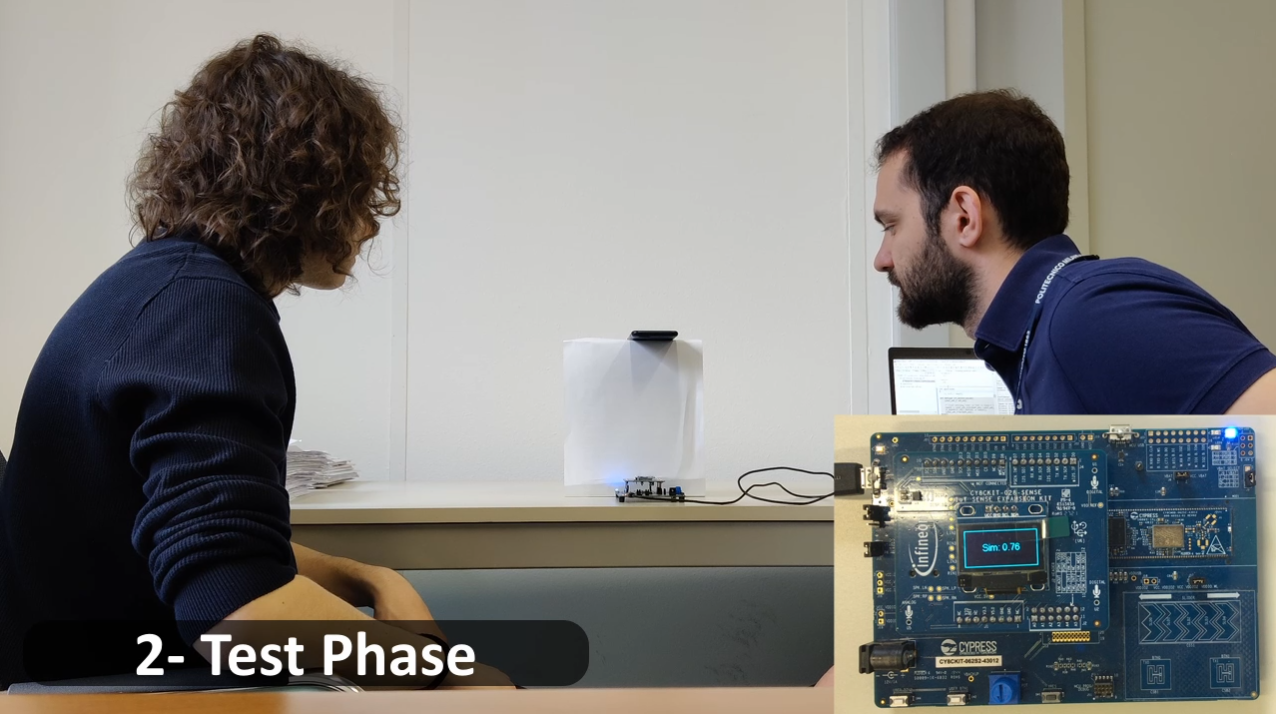}
    \caption{A frame of the video demonstrating the on-device implementation of the system.}
    \label{fig:frame}
\end{figure}

\subsection{The board}
\label{sub:board}

The considered hardware platform is the Infineon PSoC 62S2 Wi-Fi BT Pioneer Board, which is a programmable embedded system-on-chip, integrating a 150-MHz Arm® Cortex®-M4 as the primary application processor, a 100-MHz Arm Cortex-M0+ that supports low-power operations, up to 2 MB Flash and 1 MB SRAM, and the compatibility with Arduino™ shields. The application has been written to run on the Cortex®-M4 processor.
The board is also equipped with RBG LEDs, and the Infineon CY8CKIT-028-SENSE shield, which contains a digital microphone and an OLED screen.

\subsection{Implementation details}
\label{sub:impl-details}
The system has been implemented using windows of $W = 1$ s and $f_r = 16$ KHz. Each $I_t$ is consequently a 16000-element long vector. Windows are partly overlapped, and the overlapping of the window in seconds corresponds to $0.75$ s, computed as $W - T_{\Phi_k}$ where $T_{\Phi_k}$ is the inference time of $\Phi_k$. 

For the training and validation of $\Phi_k$ the Google Speech Commands dataset \cite{warden_speech_2018} has been used, while $\Phi_f$ was obtained from a model originally trained for a speaker classification task on the LibriSpeech-train-100 dataset \cite{panayotov2015librispeech}. Details on the training of $\Phi_k$ and $\Phi_C$ can be found in the project repository.

\subsection{Flash and RAM memory occupation, execution times, and power consumption}
\label{sub:measured-mem}
The on-device deployment to the board was performed through the use of the Infineon ModusToolbox \cite{Modus}, which was used also to measure the actual memory requirements on the board.
The whole application requires about 356.73 kB of flash memory to be stored. 

At runtime, the total RAM memory request is 391.92 kB. Details on the measured RAM memory occupation of each component can be found in Tab. \ref{memory-composition}.

\begin{table}[ht]
\begin{center}
\begin{tabular}{|cc|}
\hline
Component & Memory required \\ \hline \hline
$I_t$ & 32 kB     \\ \hline
$P_t$ & 7.5 kB    \\ \hline
$D_t$ & 1 kB      \\ \hline
$\Phi_k$ - weights & 104.21 kB   \\ 
$\Phi_k$ - act. & 39.68 kB   \\ \hline
$\Phi_f$ - weights & 98.08 kB  \\ 
$\Phi_f$ - act. & 70.56 kB   \\ \hline
$\Phi_c$ & 16 kB     \\ \hline
Other (application overhead) & 22.89 kB  \\ \hline
\textbf{Total}  & \textbf{391.92 kB} \\ \hline
\end{tabular}
\end{center}
\caption{Measured runtime RAM memory occupation for each component of TinySV on the PSoC 6 MCU Board.}
\label{memory-composition}
\end{table}

It's important to note that the toolbox implements a common optimization on the memory requirements for the activations of the neural networks \cite{pavan_tybox_2023}, resulting in significantly smaller memory requirements with respect to the estimation provided in Tab. \ref{parameters-composition}. 

The execution times of the two CNNs used in the application are reported in Tab. \ref{tab:time}. Compared to their execution times, the execution time of $\Phi_c$ is negligible. 

\begin{table}[ht]
\begin{center}
\begin{tabular}{|cc|}
\hline
Component & Time (s) \\ \hline \hline
$T_{MFCC}$ & 0.020 \\
$T_{\Phi_k}$ & 0.250 \\
$T_{\Phi_f}$ & 0.036 \\
$T_{\Phi_c}$ & $\sim 0$ 
 \\ \hline

\end{tabular}
\end{center}
\caption{Execution Time measured for all the modules in the on-device implementation.}
\label{tab:time}
\end{table}

While executing the application, the MCU runs at 150 MHz which is the maximum clock speed. PSoC 6 MCU operates at 3.3V. Taking into account all the active peripherals, the application consumes 19 mA of current, leading to a total power consumption of 62.7 mW. The expected runtime of the system when powered by a 1000mAh battery is 159 hours.

\section{Conclusions}
\label{conclusions}

The aim of this paper was to introduce a new type of adaptive TinyML solutions and a novel TinyML task, named TinySV, that requires the usage of on-device learning. The proposed two-layer hierarchical TinyML solution relies on two modules, i.e., Keyword Spotting and Speaker Verification, used in a cascade manner. The proposed solution adapts the TinyML model directly on-device with the data of the user, making use of a novel one-class, few-shot learning approach that deals with the lack of data and labels common to the TinyML environment. The effectiveness of the proposed solution has been successfully evaluated on a newly collected dataset that has been released to the scientific community. The efficiency of the solution has been demonstrated with the on-device implementation on an IoT device, the Infineon PSoC 62S2 Wi-Fi BT Pioneer Board, where the memory occupation, power consumption, and execution times have been evaluated. 

Future works will encompass the exploration of methods to improve the d-vector extraction, the testing of other algorithms that can be trained with a few-shot, one-class approach, and the extension of the proposed methodology to other TinyML learning tasks that have been, until now, faced only with standard supervised learning methodologies, such as object detection in pictures.

\section*{Acknowledgment}

\bibliographystyle{IEEEtran}

\bibliography{lib}

\end{document}